\documentclass[11pt,a4paper]{article} \pdfoutput=1
\usepackage{jheppub} \usepackage[T1]{fontenc}
\usepackage[english]{babel}
\usepackage{amssymb,amsthm,mathrsfs,bbm,slashed,booktabs,microtype,mathtools}
\newcommand{\dd}{\textmd{d}} \DeclareMathOperator{\Tr}{Tr}
\DeclareMathOperator{\diag}{diag}

\newcommand{\MS}{\overline{\text{MS}}} \newcommand{\OS}{\text{OS}}
\newcommand{\flow}{\text{flow}} \title{Gauge-invariant Higgs mechanism
  via gradient-flow regularization} \author{Gunnar~S.~Bali}
\affiliation{Institut für Theoretische Physik, Universität Regensburg,
  93040 Regensburg, Germany.}  \emailAdd{gunnar.bali@ur.de}
\abstract{We provide a manifestly gauge-invariant description of the
  Higgs mechanism in the Standard Model, casting it not as the
  breaking of a local symmetry, but as a smooth, analytic crossover
  from a high-temperature symmetric phase to a low-temperature Higgs
  phase. Taking the vacuum expectation value of a gauge-invariant
  local scalar operator and constructing composite quasi-particle
  states from the original fields of the Lagrangian usually introduces
  contact-term divergences. Previous attempts to circumvent this issue
  via bilocal operators remain unsatisfactory. Here we demonstrate
  that the gradient flow at a flow time $t$ provides a clean,
  structurally sound, and Lorentz-covariant regularization that
  permits a consistent, renormalon-free matching onto the standard
  $\overline{\text{MS}}$ scheme at a scale $\mu \sim 1/\sqrt{2t}\gg
  m_t$. Beyond its theoretical appeal, this construction clarifies a
  number of conceptual and practical questions. We demonstrate the
  utility of this framework at the one-loop level, thereby
  establishing the foundation for future electroweak precision physics
  in this approach.}

\keywords{Gauge Symmetry, Renormalization and Regularization, Renormalons, Higgs Properties, Electroweak Precision Physics.}
\arxivnumber{2609.12877}
\begin{document}
\maketitle
\section{Introduction}
The usual textbook description of the Standard Model of elementary
particles and their interactions relies heavily on the concept of
electroweak ``symmetry breaking'' (EWSB). In this framework, the
scalar Higgs field acquires a non-vanishing vacuum expectation value
(VEV), $v$, which acts as an order parameter that ``breaks'' the local
$SU(2)_L \times U(1)_Y$ gauge symmetry down to $U(1)_{\text{em}}$ and
endows three vector bosons and the fermions with mass. While this
narrative constitutes an exceptionally successful tool in unitary
gauge at tree-level and in 't Hooft
$R_\xi$-gauges~\cite{tHooft:1971qjg} for higher-order perturbation
theory, it obscures the underlying quantum field theory. In
particular, Elitzur's theorem~\cite{Elitzur:1975im,DeAngelis:1977su}
strictly prohibits the spontaneous breaking of a local gauge symmetry.
This is because gauge redundancies do not constitute symmetries of
physical observables.

One would aim at a rigorous, non-perturbative resolution of this
tension. While a fully consistent formulation of chiral gauge theories
in lattice regularization is still an open question, there has been
substantial progress, see,
e.g.~\cite{Grabowska:2016bis,Clancy:2023ino,Golterman:2025boq} and
references therein. In lattice regularization the Higgs
mechanism~\cite{Anderson:1963pc,Englert:1964et,Higgs:1964pj,Guralnik:1964eu,Kibble:1967sv}
(reviewed, e.g., in~\cite{Maas:2017wzi}) can be understood entirely
through gauge-invariant variables.  Non-perturbative
simulations~\cite{Kajantie:1995kf,Kajantie:1996mn,Karsch:1996yh,Aoki:1996cu,Csikor:1998eu}
demonstrate that the transition (around a temperature
$T_{\text{EWSB}}\approx
160\,$GeV~\cite{DOnofrio:2014rug,DOnofrio:2015gop,Gould:2022ran}) from
the symmetric phase to the low-temperature Higgs phase constitutes a
smooth, analytic crossover rather than a thermodynamic phase
transition. The non-perturbative formulation clarifies that also in
the symmetric phase the expectation value of the Higgs condensate does
not vanish, due to thermal fluctuations and persistent interactions.
No local symmetry is broken; instead, the physical degrees of freedom
rearrange themselves across the two regimes.

Mapping this gauge-invariant description onto the elementary fields of
the Lagrangian is not straightforward. In standard continuum field
theory, identifying a gauge-invariant low-temperature Higgs
quasi-particle that has a well-defined VEV is severely hampered by
power-law divergences related to contact terms. Fröhlich, Morchio, and
Strocchi (FMS)~\cite{Frohlich:1980gj,Frohlich:1981yi} introduced a
bilocal operator framework to circumvent this problem, but their
reliance on spatially separated fields introduces undesirable
non-localities and remains phenomenologically unsatisfactory. More
recently, this picture has been investigated by Maas and
Sondenheimer~\cite{Maas:2020kda} at the one-loop level. A related
approach is the dressing field
method~\cite{Fournel:2012cr,Francois:2017akk,Francois:2024rdm}.

In this paper, we demonstrate that the gradient
flow~\cite{Narayanan:2006rf,Luscher:2009eq,Luscher:2010iy,Luscher:2011bx}
solves this structural bottleneck. The gradient flow acts as a
manifestly Lorentz-covariant and gauge-invariant regulator. The
corresponding heat kernel smooths out fluctuations at length scales
smaller than $\sqrt{8t}$, where $t$ is the flow time, regularizing
ultraviolet (UV) divergences of composite operators in a structurally
clean way, before perturbative expansions are executed. As a hard
cutoff, the gradient flow is sensitive to power divergences, which has
several structural advantages. Moreover, unlike dimensional
regularization, the gradient flow is defined beyond perturbation
theory.

The suggested framework is not only structurally and conceptually
appealing, but it may also prove highly advantageous for precision
electroweak phenomenology. By shielding the renormalization group
trajectories of the short-distance couplings from infrared (IR)
renormalons that plague on-shell mass and VEV definitions, it provides
a mathematically stable coordinate system for high-order calculations.
Crucially, the resulting picture can be matched in a consistent,
renormalon-free way onto conventional continuum subtraction schemes
like the $\overline{\text{MS}}$ scheme. The matching itself does not
require the theory to be defined at the singular $t \to 0^+$ boundary,
provided that the matching scale $\mu\sim 1/\sqrt{2t}$ is chosen within
a range in which perturbation theory is applicable. We demonstrate the
practical utility of this architecture by carrying out the matching
explicitly at the one-loop level, providing a baseline for future
precision Standard Model calculations.

The gradient flow was originally introduced in the context of the
non-perturbative evaluation of observables in lattice-regularized
non-Abelian gauge theories. Its application to the electroweak sector
of the Standard Model addresses a fundamentally different question.
The Higgs-Yukawa sector is inherently perturbative; however, if a
cutoff scale like the Planck mass is introduced or heavy particles
are added, the theory becomes highly sensitive to small variations in
the UV, because in this case the Higgs mass is not protected by any
symmetry. In this context, the gradient flow provides a clean
formalism for scale separation, shielding low-energy aspects from new
physics at high scales without distorting any fundamental properties
of the theory. Although we argue that the Higgs mechanism should be
formulated using a non-perturbative framework, we do not advocate
simulating the full Standard Model directly on the lattice. Since the
electroweak symmetry-breaking scale satisfies $v \sim
1000\,\Lambda_{\text{QCD}}$ and the interaction of the top quark with
gluons cannot be neglected, a naive simulation of the Standard Model
is not feasible with present numerical methods; due to this large
separation of scales, the lattice volume would have to be
enormous. This can of course be mitigated to some extent by effective
field theory (EFT) approaches, for instance exploiting dimensional
reduction at large
temperature~\cite{Appelquist:1981vg,Bali:1993tz,Farakos:1994xh}.

At a matching scale $\mu\approx 2m_t$, the values of the largest
renormalized Standard Model couplings read $\alpha_s(\mu) =
g_s^2(\mu)/(4\pi) \approx 0.10$ and $y_t^2(\mu)/(4\pi) \approx
0.07$~\cite{Buttazzo:2013uya}. At these short distances, perturbation
theory entirely suffices to match the fields and couplings of the
symmetric Lagrangian to the theory defined by the relevant degrees of
freedom in the Higgs phase (the so-called physical basis). In fact,
all short-distance Standard Model phenomenology is amenable to
perturbation theory. Nevertheless, establishing a rigorous,
gauge-invariant non-perturbative definition of the Standard Model and
EWSB provides the conceptual clarity that is necessary to protect
against ambiguities at high-loop orders.  We remark that in spite of
the applicability of perturbation theory for short-distance physics,
even at high temperatures magnetic screening and topological sphaleron
transitions keep the long-distance sector fundamentally
non-perturbative.

This article is organized as follows. In section~\ref{sec:standard},
we outline approaches to address EWSB in the Standard Model, introduce
our notation and define the relevant gauge-invariant composite
operators. We then define the gradient flow for all Standard Model
fields and discuss the associated renormalization in
section~\ref{sec:gradient}, where we also provide some one-loop
renormalization constants. In section~\ref{sec:match}, we discuss and
carry out the matching between gradient-flow and dimensional
regularization in the modified minimal subtraction ($\MS$) scheme. We
pay particular attention to the matching of the scalar bilinear
operator, which undergoes mixing with the identity, and suggest a
procedure that ensures the cancellation of the leading renormalon
contribution. In section~\ref{sec:cross}, we connect back to the
foundational concepts by comparing the breaking of the approximate
chiral symmetry of QCD with the crossover to the low-temperature Higgs
phase of the Standard Model, before we summarize.

\section{The Standard Model at high and low temperature}
\label{sec:standard}
\subsection{The FMS mechanism in a nutshell}
In the low-temperature Higgs phase of the Standard Model, the fundamental
scalar doublet field
\begin{align}
  \label{eq:higgs1}
  \Phi(x)=\frac{1}{\sqrt{2}}\left(\begin{array}{c}\phi_2(x)+i\phi_1(x)\\
    \phi_0(x)-i\phi_3(x)\end{array}\right)
\end{align}
fluctuates around a non-zero classical minimum value
$v^2=2\Phi_0^\dagger\Phi_0=m^2/\lambda$ of the Higgs potential
\begin{align}
  \label{eq:higgs}
  V(\Phi)=-m^2\Phi^{\dagger}\Phi+\lambda(\Phi^{\dagger}\Phi)^2,
\end{align}
where $v \approx 246\,\text{GeV}$
and
\begin{align}
  \Phi_0(x)=U(x)\Phi_0,\quad \Phi_0=\frac{v}{\sqrt{2}}\left(\begin{array}{c}0\\1\end{array}\right),\quad U(x)\in SU(2).\label{eq:ux}
\end{align}
For convenience we also introduce the conjugate field
\begin{align}
  \Phi^c(x)=i\sigma_2\Phi^*(x)=\frac{1}{\sqrt{2}}\left(\begin{array}{c}\phi_0(x)+i\phi_3(x)\\-\phi_2(x)+i\phi_1(x)\end{array}\right)
\end{align}
and the matrix-valued scalar field
\begin{align}
  \label{eq:lin}
  \boldsymbol{\Phi}(x)=\left(\Phi^c(x),\Phi(x)\right)
  =\frac{1}{\sqrt{2}}\left(\phi_0(x)\mathbbm{1}+i\vec{\phi}(x)\cdot\vec{\sigma}\right).
\end{align}

In standard introductory presentations of EWSB in the Standard Model
(see, e.g.~\cite{Donoghue:1992dd,Denner:2019vbn}), a gauge is chosen such that
fluctuations $\phi^+(x)=[\phi_2(x)+i\phi_1(x)]/\sqrt{2}$,
$\chi(x)=-\phi_3(x)$ and
$h(x)\in\mathbb{R}$ around $\Phi_0(x)=\Phi_0$ are small:
\begin{align}
  \Phi(x)=\frac{1}{\sqrt{2}}\left(\begin{array}{c}\sqrt{2}\phi^+(x)\\v+h(x)+i\chi(x)\end{array}\right).
\end{align}
In unitary gauge ($\phi^+=\chi=0$) only the lower component
$(v+h(x))/\sqrt{2}$ is non-trivial. In this case, the physical
particle content takes a particularly simple form and the Higgs
mechanism can be interpreted as spontaneous symmetry breaking with the
would-be Goldstone bosons being ``eaten'' by three vector bosons that
become massive.  In contrast, in a gauge-covariant formulation $U(x)$
in eq.~\eqref{eq:ux} is non-trivial and the expectation value of the
Higgs doublet must vanish identically: $\langle \Phi(x) \rangle =
(0,0)^\intercal$.

To preserve gauge covariance, the FMS
framework~\cite{Frohlich:1980gj,Frohlich:1981yi} reinterprets the
physical particle spectrum in terms of gauge-invariant composite bound
states, see
also~\cite{Fournel:2012cr,Francois:2017akk,Maas:2017wzi,Maas:2020kda,Creutz:2023wxu,Maas:2023emb,Francois:2024rdm}. For
example, one would like to relate the Higgs boson $h(x)$ to the
difference between the gauge-invariant local singlet operator $2O_H(x)
= 2\Phi^\dagger(x)\Phi(x)$ and its value $v^2$ at the minimum of the
Higgs potential. However, at the quantum level, a severe structural
obstacle is encountered: evaluating products of field operators at the
same spacetime point $x$ triggers UV contact-term divergences
proportional to the square of the UV cutoff. Therefore, in the FMS
framework local operators are abandoned in favour of bilocal operators
connected by Wilson lines $U(y,x)$:
\begin{equation}
\mathcal{O}(x,y) = \Tr\left[ \boldsymbol{\Phi}^\dagger(y) \, U(y,x) \, \boldsymbol{\Phi}(x) \right]=2\,\Re\left[\Phi^\dagger(y)\,U(y,x)\,\Phi(x)\right].
\end{equation}
By taking the equal-time infinite-separation limit
($\lim_{|x-y|\to\infty} \langle \mathcal{O}(x,y) \rangle = v^2$,
setting aside the renormalization of the Wilson line), the FMS
framework successfully evades the zero-distance singularity. However,
the reliance on bilocal tracking parameters complicates the
mathematical formulation and prevents a straightforward physical
interpretation of the quasi-particles of the broken phase as
point-like composite states at the electroweak scale.

In this paper, we demonstrate that gauge-covariant gradient flow
provides an elegant solution to this long-standing dilemma. By
evolving the fundamental fields along a fictitious extra
dimension --- parameterized by the flow time $t$ --- short-distance
quantum fluctuations are smoothed out in the $t>0$ bulk over a
Euclidean distance $\sqrt{8t}$. At any positive flow time, this
damping regularizes the contact-term singularities, rendering the
flowed local operator
\begin{align}
  O_H(x;t)=\Phi^\dagger(x;t)\Phi(x;t)
\end{align}
UV-finite and enabling a physically clean, conceptually sound and
manifestly gauge-invariant framework of EWSB.

\subsection{Standard approaches to EWSB}
The Standard Model is an EFT that is only valid up to a cutoff scale
$\Lambda$, due to the Landau poles of the $U(1)_Y$ and Yukawa
couplings. However, modern
analyses~\cite{Bezrukov:2012sa,Degrassi:2012ry,Buttazzo:2013uya,Bednyakov:2015sca,Bednyakov:2025uur}
indicate that the cutoff could well be as large as the Planck mass.
The Standard Model may also be embedded in an asymptotically safe way
into grand unified theories, e.g.\ $SU(2)_L\times SU(4)\times
SU(2)_R\subset SO(10)$, see~\cite{Molinaro:2018kjz} and references
therein.

The Standard Model (including right-handed Weyl neutrinos) contains 20
independent couplings in the Yukawa sector --- 10 in the quark and 10 in
the lepton sector --- as well as the $U(1)_Y$, $SU(2)_L$ and $SU(3)_C$
couplings $g_1=g'$, $g_2=g$ and $g_3=g_s$, and the quartic Higgs
coupling $\lambda$. All these couplings are dimensionless. Only one
term of the Lagrangian, $m^2\Phi^\dagger\Phi$, is relevant and the
accompanying quadratic Higgs coupling $m^2$ provides the scale. In
dimensional regularization all couplings including $m^2$ only
renormalize multiplicatively.  As in QED, for Standard Model
perturbative expansions carried out in the $\MS$ scheme no cutoff is
needed. In particular, the mass parameter $m$ is protected against
additive renormalization by dimensional counting and classical scale
symmetry: if $m(\mu)=0$, it remains zero under renormalization. Large
additive counter-terms $\sim\Lambda^2$ are however encountered if a
cutoff scale is introduced or if particles with masses much bigger
than the electroweak scale are added, giving rise to the so-called
naturalness problem.

The Lagrangian and couplings are defined in the symmetric
high-temperature phase. However, experiment takes place in the
low-temperature domain, where the effective degrees of freedom differ
from the fundamental fields of the Lagrangian. For instance, the
$SU(2)_L$ tree-level coupling can in principle be obtained in this
Higgs phase from the on-shell mass of the $W$ boson via the relation
$g^2(\mu)=4M_W^2/v^2(\mu)$. Higher-order corrections to this relation
depend on the scheme. Also the relation
$v^2(\mu)=m^2(\mu)/\lambda(\mu)$ between the position of the minimum
of the Higgs potential and the quadratic and quartic Higgs couplings
encounters corrections in gauge-dependent schemes, where an effective
potential $V_{\text{eff}}(h)$ is introduced. In particular, the VEV of
the real scalar Higgs field $h(x)$ receives additive contributions
from tadpole (sub-)diagrams. The resulting large corrections are
mainly addressed in two ways.

In the Fleischer-Jegerlehner (FJ) scheme~\cite{Fleischer:1980ub} the
tadpole contributions to $h(x)$ are absorbed into the counter terms
and field redefinitions. The disadvantage is that the expansion is
carried out about the tree-level vacuum parameter $v$, which does not
correspond to the minimum of the renormalized effective Higgs
potential anymore, resulting in poor perturbative convergence for
certain calculations. In the tadpole-free scheme (also dubbed
parameter renormalized tadpole scheme,
PRTS)~\cite{Denner:1991kt,Martin:2014cxa}, the vacuum parameter $v$
always tracks the minimum of the full renormalized potential such that
tadpole contributions to $h(x)$ vanish. The relation between the bare
and renormalized couplings is modified accordingly. While the FJ
scheme is gauge-covariant, PRTS results for gauge-invariant operators
and Green functions generally exhibit a gauge-dependence, though this
dependence is expected to reduce at higher perturbative
orders. Recently, Dittmaier and Rzehak~\cite{Dittmaier:2022maf}
suggested a new gauge-invariant VEV scheme (GIVS) that combines the
benefits of both choices --- namely, gauge independence and expansion
about the true minimum of the renormalized Higgs potential --- by
switching to a non-linear parametrization of the Higgs doublet.

The choice of scheme fundamentally affects both the notation and the
matching procedure. A key advantage of the PRTS is that the relation
$v(\mu)=m(\mu)/\sqrt{\lambda(\mu)}$ is maintained at all orders of
perturbation theory. In our context a gauge-independent scheme that
preserves this exact property is necessary. We explain our
conventions, which are conceptually close to the GIVS, in the
following subsection.

\subsection{Gauge-invariant formulation of EWSB}
In standard perturbative treatments of the electroweak sector in the
FJ~\cite{Fleischer:1980ub} or PRTS
schemes~\cite{Denner:1991kt,Martin:2014cxa} the introduction of a
linear field shift about a classical expectation value inherently
introduces gauge-dependent configurations. As a consequence, loop
corrections contaminate the vacuum structure, causing the physical
minimum of the effective Higgs potential to depart from the ratio of the
renormalized Lagrangian parameters, $v^2(\mu) =
m^2(\mu)/\lambda(\mu)$.

In this work, we adopt an alternative, mathematically rigorous
convention by maintaining strict local gauge invariance throughout the
renormalization and matching procedures. By avoiding gauge-variant
coordinate choices, loop corrections to the Higgs sector from the
gauge, scalar, and heavy fermion interactions are absorbed into the
renormalization group running of the fundamental couplings, $m(\mu)$
and $\lambda(\mu)$, preserving the tree-level relation $v^2(\mu) =
m^2(\mu)/\lambda(\mu)$ to all orders, where $v^2(\mu)$ is not the VEV
but tracks the minimum of the potential.  All fields of the
low-temperature Lagrangian are then defined as fluctuating about the
critical gauge orbit $\Phi^\dagger\Phi=v^2/2$ and not about a VEV that
changes with the order of perturbation theory.

The physical, off-shell, and momentum-dependent corrections are mapped
directly onto composite operators within a systematic operator product
expansion (OPE). The emergence of particle masses in the
low-temperature phase is understood not as the spontaneous breaking of
a local symmetry, but as a consequence of the dimensionful coupling
$m$ and the shape of the Higgs potential. The only exact global
symmetry of the Standard Model (with right-handed Weyl neutrinos) in
both phases is $U(1)_{B-L}$. The gradient-flow regulator ensures a
clean separation of scales, shielding the low-energy observables from
unphysical coordinate artefacts and UV divergences, while leaving the
underlying symmetries of the theory intact.

Like in GIVS~\cite{Dittmaier:2022maf}, our starting point is the
non-linear parametrization~\cite{Kibble:1967sv}
\begin{align}
  \boldsymbol{\Phi}(x)
  =\frac{\rho(x)}{\sqrt{2}}e^{i\vec{\zeta}(x)\cdot\vec{\sigma}},\quad \rho(x)=v+h(x)
\end{align}
of eq.~\eqref{eq:lin}, where, defining
$\phi=(\vec{\phi}\cdot\vec{\phi})^{1/2}$ and
$\zeta=(\vec{\zeta}\cdot\vec{\zeta})^{1/2}$,
$\rho^2=(\phi_0^2+\phi^2)$, $\phi_0=\rho\cos\zeta$ and
$\vec{\phi}=\rho\,\vec{\zeta}\,\zeta^{-1}\!\sin\zeta$.  Then only the
field components $\zeta_i(x)$ are subject to gauge
transformations. Note that this corresponds to the matrix form of
$\Phi_0(x)$ of eq.~\eqref{eq:ux} with the substitution
$v\mapsto\rho(x)$ and $U=e^{i\vec{\zeta}(x)\cdot\vec{\sigma}}$. In
this parametrization, the Higgs doublet reads (see
eq.~\eqref{eq:higgs1})
\begin{align}
  \Phi(x)=\frac{1}{\sqrt{2}}(v+h(x))
  \begin{pmatrix}\zeta_2(x)s_\zeta(x)+i\zeta_1(x)s_\zeta(x)\\
    c_\zeta(x)-i\zeta_3(x)s_\zeta(x)\end{pmatrix},\quad
  c_\zeta\coloneqq\cos\zeta,\quad s_\zeta\coloneqq
  \frac{\sin\zeta}{\zeta}.
\end{align}
Gauge-transforming $\Phi'=U_L\Phi e^{i\varphi}$ with $U_L\in SU(2)_L$,
results in
\begin{align}
  \boldsymbol{\Phi}'=U_L
  \boldsymbol{\Phi}R^\dagger,\quad R=e^{i\sigma_3\varphi}\in SU(2)
\end{align}
for the matrix representation $\boldsymbol{\Phi}=(i\sigma_2\Phi^*,\Phi)$,
re-confirming the gauge-invariance of $\rho(x)$. 

Since $h(x)=\rho(x)-v$ is manifestly gauge-invariant, the relation
\begin{align}
  \left\langle O_H(x;t)\right\rangle=\frac12
  \left\langle\Tr[\boldsymbol{\Phi}^\dagger(x;t)\boldsymbol{\Phi}(x;t)]\right\rangle
  =\frac12\left\langle\rho^2(t)\right\rangle=\frac{v^2(t)}{2}+
  v(t)\left\langle h(t)\right\rangle
  +\frac12\left\langle h^2(t)\right\rangle
\end{align}
is gauge-independent and exact for flow times $t>0$. Since the
expectation value is translationally invariant, we have dropped the
$x$-dependence. Also, there is an additive $1/t$-dependence, see
eq.~\eqref{eq:ope1} below, which must be subtracted.

\subsection{Construction of gauge-invariant composite fields}
\begin{table}[tbp]
\centering
\caption{Composite operators that are invariant under gauge
  transformations within the screened coset space, at a flow time
  $t>0$.  The right-handed fermions (except for neutrinos) interact
  electromagnetically or strongly. $D_\mu(t)$ is the covariant
  derivative, defined in eq.~\eqref{eq:cov} (for $Y=1/2$, $\delta_L=1$
  and $\delta_C=0$). The $\mathbf{W}^1$, $\mathbf{W}^2$ and
  $\mathbf{W}^3$ combinations combine into the $W^{\pm}$ and $Z^0$
  bosons according to eq.~\eqref{eq:boson}.  In the last column, we
  display the engineering mass dimensions.}
\label{tab:operators}
\begin{tabular}{llc}
\toprule
Degree of freedom & Operator &Dimension \\
\midrule
Scalar Higgs & $O_H(x;t) = \Phi^\dagger(x;t)\Phi(x;t)=\frac12\!\Tr[\boldsymbol{\Phi}^\dagger(x;t)\boldsymbol{\Phi}(x;t)]$ & $2$\\
\addlinespace
Left-handed fermion & $\Psi^{u,\nu}_{L}(x;t) = \frac{\sqrt{2}}{v(t)} \Phi^{c\dagger}(x;t) \psi_L^{u,\nu}(x;t)$ & $3/2$\\
\addlinespace
Left-handed fermion & $\Psi^{d,e}_{L}(x;t) = \frac{\sqrt{2}}{v(t)} \Phi^\dagger(x;t) \psi_L^{d,e}(x;t)$ & $3/2$\\
\addlinespace
Right-handed fermion & $\Psi^{u,\nu,d,e}_{R}(x;t) = \psi_R^{u,\nu,d,e}(x;t)$ & $3/2$\\
\addlinespace
Fermion mass term & $O^{u,\nu,d,e}_{F,ij}(x;t) = \bar{\Psi}^{u,\nu,d,e}_{L,i} \Psi^{u,\nu,d,e}_{R,j} + \bar{\Psi}^{u,\nu,d,e}_{R,i} \Psi_{L,j}^{u,\nu,d,e}$ & $3$\\
\addlinespace
Massive bosons& $\mathbf{W}^a_\mu(x;t) = -\frac{i}{2g v^2(t)}\! \Tr \left[\boldsymbol{\Phi}^\dagger(x;t) \overleftrightarrow{D}_\mu(t) \boldsymbol{\Phi}(x;t)\sigma^a \right]$ & $1$\\
\addlinespace
Photon field & $A_\mu(x;t) = \cos\theta_W B_{\mu}(x;t) + \sin\theta_W \mathbf{W}_{\mu}^3(x;t)$& $1$\\\addlinespace
Gluon fields &$G^A_\mu(x;t)$&$1$\\
\bottomrule
\end{tabular}
\end{table}

The Standard Model Lagrangian, expressed in terms of both fundamental
fields and fields fluctuating about the minimum of the Higgs
potential, can be found in~\cite{Donoghue:1992dd}. We do not replicate
this here. We start with the field content --- the complex Higgs doublet
$\Phi$, the left- and right-handed Weyl fermions $\psi_L$ and
$\psi_R$, the $U(1)_Y$ gauge field $B_\mu$, the $SU(2)_L$ gauge fields
$W_{\mu}^a$ ($a=1,2,3$) and the $SU(3)_C$ gauge fields $G_\mu^A$
($A=1,\ldots,8$) --- to construct composite operators at a gradient flow
time $t$. These operators are constructed to be invariant under local
gauge transformations within the $[SU(2)_L\times
  U(1)_Y]/U(1)_{\text{em}}$ coset space. On this $S^3$ space an
$SU(2)$ group structure is natively realized. The corresponding gauge
freedom becomes screened when $O_H(x;t)$ acquires a non-vanishing
background value $v^2(t)/2$ at low temperatures, aligning itself with
the minimum of the Higgs potential. Specifically, this screening acts
on the $SU(2)_L$ generators $T^1$ and $T^2$, alongside a distinct
linear combination of the diagonal $T^3$ component with the
hypercharge generator. Gauge invariance then locks the remaining
active degrees of freedom to the corresponding gauge orbit of the
scalar doublet field.

The $SU(2)$ gauge-singlet operators that can be constructed in this
way are listed in table~\ref{tab:operators}. These constitute the
fields of the low-temperature Lagrangian (physical basis), with the
exception of $O^{u,\nu,d,e}_{F,ij}$. Note that the latter fermion mass
terms are $U(1)_{\text{em}}$ singlets ($Q=Y+T_3=0$) and $SU(3)_C$
singlets.  Rewriting the original Lagrangian in terms of the screened
operators, these terms will be accompanied by $v(t)/\sqrt{2}$ times
the Yukawa matrices $Y^u_{ij}$, $Y^d_{ij}$, $Y^\nu_{ij}$ and
$Y^e_{ij}$, where the $Y^d$- and $Y^e$-couplings accompany composite
left-handed fermions, constructed with $\Phi$ to give $T_3=0$, and the
$Y^u$- and $Y^{\nu}$-couplings accompany fermions, constructed with
$\Phi^c$. Note that the eigenvalues of $Y^uY^{u\dagger}$ are the
squared Yukawa couplings $y^2_u$, $y^2_c$ and $y^2_t$ and the
eigenvalues of $Y^{\nu}Y^{\nu\dagger}$ are $y^2_1$, $y^2_2$ and
$y^2_3$, corresponding to the neutrino masses $m_i=y_iv/\sqrt{2}$,
etc.. A standard general non-redundant parametrization of the Yukawa
matrices (see, e.g.~\cite{Donoghue:1992dd}) reads
$Y^u=\diag(y_u,y_c,y_t)$, $Y^d=V\cdot\diag(y_d,y_s,y_b)$,
$Y^{\nu}=U\cdot\diag(y_1,y_2,y_3)$ and $Y^e=\diag(y_e,y_\mu,y_\tau)$
with the unitary Cabibbo-Kobayashi-Maskawa (CKM) and
Pontecorvo-Maki–Nakagawa–Sakata (PMNS) matrices $V$ and $U$,
respectively, that each contain four independent parameters.  In the
low-temperature theory, a left-handed composite and a right-handed
fundamental Weyl spinor can be combined into a non-chiral Dirac spinor
$\Psi$. In the chiral basis with $\gamma_5=\diag(-1,-1,1,1)$, the
latter is given as $\Psi=(\Psi_L,\Psi_R)^\intercal$, and its Dirac
adjoint as $\bar{\Psi}=(\bar{\Psi}_R,\bar{\Psi}_L)$. Then
$O_F=\bar{\Psi}\Psi$.

In order to account for the backwards gradient-flow evolution for
the  adjoint particles, in the table we use the symmetric derivative
\begin{align}
  \label{eq:symderiv}
  \overleftrightarrow{\slashed{D}}=\slashed{D}-\overleftarrow{\slashed{D}},\quad
  \slashed{D}=D_{\mu}\sigma_{\mu},\quad \sigma_0=\mathbbm{1},
\end{align}
where $\chi\overleftarrow{D}\coloneqq (D\chi^{\dagger})^\dagger$ and
the $\sigma_\mu$ above are Pauli matrices acting on Weyl spinor space,
not on $SU(2)_L$. The physical massive vector bosons are obtained as
\begin{align}
  \label{eq:boson}
  W^{\pm}_\mu=\frac{1}{\sqrt{2}}(\mathbf{W}^1_\mu\mp
  i\mathbf{W}^2_\mu), \quad Z^0_\mu=\cos\theta_W
  \mathbf{W}_\mu^3-\sin\theta_W B_\mu,
\end{align}
with the tree-level relations $\tan\theta_W=g'/g$, $M_W=g\,v/2$ and
$M_Z=\sqrt{g^2+g^{\prime\,2}}\,v/2$, where $\theta_W$ is the Weinberg
angle. The covariant derivative at a flow time $t$ reads
\begin{equation}\label{eq:cov}
D_\mu(t) = \left(\partial_\mu + i g' B_\mu(x;t) Y\right)\mathbbm{1} + i g\, \delta_L W_\mu^a(x;t) T^a + i g_s\delta_C G_{\mu}^A(x;t)t^A,
\end{equation}
where $\delta_L,\delta_C\in\{0,1\}$, depending on whether the argument
is a singlet or a non-singlet under $SU(2)_L$ and $SU(3)_C$. In the
fundamental representation, $T^a=\sigma^a/2$ and $t^A=\lambda^A/2$.
For the Higgs field the weak isospin is $Y=1/2$, $\delta_L=1$ and
$\delta_C=0$. $D_\mu=D_{\mu}(0)$ is the standard covariant derivative
at the boundary. The above conventions are suitable for perturbative
expansion. For non-perturbative aspects, e.g.\ defining the
gradient-flow or carrying out lattice computations, it is convenient
to absorb the couplings $g'$, $g$ and $g_s$ into the definition of the
gauge fields.

The low-temperature Lagrangian is identical to the Lagrangian in the
symmetric phase, but re-expressed in terms of the composite fields
that are relevant near the minimum of the Higgs potential. No
higher-dimensional operators are introduced and all terms of the
Lagrangian remain marginal, with the exception of the quadratic Higgs
term: the Standard Model in the symmetric and in the Higgs phase is
the same. In particular, the latter does not constitute a low-energy
EFT description of the former. Note, however, that higher-dimensional
operators can appear in the matching between the gradient-flow and the
$\MS$ schemes and in the matching between the fields of the two
Lagrangians. This is also the case for the OPE of bilocal FMS
operators. Overall, in the low-temperature theory, which is locked to
the critical gauge orbit, the Higgs degrees of freedom are reduced
from four to one. At the same time, the three composite vector bosons
have become massive, meaning that each of them has three on-shell
polarizations rather than the two in the massless case. The counting
in this picture is the same as in the gauge-fixed approach.

The real scalar Higgs field of the low-temperature phase is related to
fluctuations of $O_H(x;t)$ about $v^2(t)/2$.  However, $O_H$ will mix
with the unity operator, which at the boundary ($t=0$) is the source
of the power divergence of its vacuum expectation value. This
divergence is avoided both by the FMS construction and the gradient
flow. We will see that in the latter case this term can safely be
identified and subtracted.

\section{Gradient flow of the Standard Model}
\label{sec:gradient}
\subsection{The flow equations}
Following the considerations of~\cite{Luscher:2013cpa}, we formulate
the gradient-flow equations strictly within $d$-dimensional Euclidean
spacetime (where $d=4-2\epsilon$), thereby ensuring that the diffusion
operator remains elliptic and well-posed. Physical observables in
Minkowski spacetime are subsequently recovered via standard analytic
continuation. The gauge-covariant gradient flow of the scalar doublet
is defined via
\begin{align}
  \frac{\partial \Phi(x;t)}{\partial t} &= D_\mu(t) D_\mu(t) \Phi(x;t),\quad
  \Phi(x,0)=\Phi(x),\\
  \frac{\partial \Phi^\dagger(x;t)}{\partial t} &= \Phi^{\dagger}(x;t)\overleftarrow{D}_\mu(t) \overleftarrow{D}_\mu(t),\quad
  \Phi^\dagger(x,0)=\Phi^\dagger(x).
\end{align}
We will follow the anti-Hermitian conventions of the
literature~\cite{Luscher:2010iy} for the fields in the flow-time
evolution. The covariant derivative~\eqref{eq:cov} then reads
\begin{align}
  D_\mu(t) &= \left(\partial_\mu + \mathcal{B}_\mu(x;t) Y\right)\mathbbm{1} + \delta_L \mathcal{W}_\mu(x;t) + \delta_C\, \mathcal{G}_{\mu}(x;t),\\
  \mathcal{B}_\mu(x;t)&=ig'B_\mu(x;t),\quad
  \mathcal{W}_\mu(x;t)=igW^a_\mu(x;t)T^a,\quad
  \mathcal{G}_\mu(x;t)=ig_sG^A_\mu(x;t)t^A.
\end{align}
Note that in the latter two cases, in the gradient-flow literature the
factor $i$ is absorbed into the definition of the generators
$\mathcal{T}^a=-iT^a$. Then
$[\mathcal{T}^a,\mathcal{T}^b]=f_{abc}\mathcal{T}^c$,
$\mathcal{W}^a_\mu=-gW^a_\mu$ and $\mathcal{G}^A_\mu=-g_sG^A_\mu$.

The gauge connections themselves evolve along their corresponding
non-Abelian and Abelian gradient flow trajectories~\cite{Luscher:2010iy}
\begin{align}
  \frac{\partial \mathcal{W}_\mu(x;t)}{\partial t} &= D_\nu(t) \mathcal{W}_{\nu\mu}(x;t), \quad
  \mathcal{W}_\mu(x,0)=\mathcal{W}_{\mu}(x),\quad
  \mathcal{W}_{\mu\nu}=\left.[D_{\mu},D_{\nu}]\right|_{\begin{subarray}{l}\delta_L=1\\Y=\delta_C=0\end{subarray}},\label{eq:su2flow}\\
  \frac{\partial \mathcal{G}_\mu(x;t)}{\partial t} &= D_\nu(t) \mathcal{G}_{\nu\mu}(x;t), \quad
  \mathcal{G}_\mu(x,0)=\mathcal{G}_{\mu}(x),\quad
  \mathcal{G}_{\mu\nu}=\left.[D_{\mu},D_{\nu}]\right|_{\begin{subarray}{l}\delta_C=1\\Y=\delta_L=0\end{subarray}},\\
  \frac{\partial \mathcal{B}_\mu(x;t)}{\partial t} &= \partial_\nu \mathcal{B}_{\nu\mu}(x;t),
  \quad \mathcal{B}_{\mu}(x,0)=\mathcal{B}_{\mu}(x),\quad \mathcal{B}_{\mu\nu}=\partial_{\mu}\mathcal{B}_\nu-\partial_\nu \mathcal{B}_\mu,
\end{align}
where, for example when $Y=\delta_C=0$, the connection reduces to
$D_\mu=\partial_\mu+[\mathcal{W}_{\mu},\cdot]$. Furthermore,
$\mathcal{W}_{\mu\nu}=igW_{\mu\nu}$,
$\mathcal{G}_{\mu\nu}=ig_sG_{\mu\nu}$
and
$\mathcal{B}_{\mu\nu}=ig'B_{\mu\nu}$.
Finally, the fundamental Weyl fermions
flow according to~\cite{Luscher:2013cpa}
\begin{align}
  \frac{\partial\psi_{L,R}(x;t)}{\partial t}&=D_\mu(t) D_\mu(t)\,\psi_{L,R}(x;t),\quad \psi_{L,R}(x,0)=\psi_{L,R}(x),\\
  \frac{\partial\bar{\psi}_{L,R}(x;t)}{\partial t}&=\bar{\psi}_{L,R}(x;t)\,\overleftarrow{D}_\mu(t) \overleftarrow{D}_\mu(t),\quad\bar{\psi}_{L,R}(x,0)=\bar{\psi}_{L,R}(x),
\end{align}
where the covariant derivative depends on the chirality ($\delta_L=1$
for left-handed and $\delta_L=0$ for right-handed fermions), on
whether the fermion is a quark ($\delta_C=1$) or a lepton
($\delta_C=0$) and the weak hypercharge $Y=Q-T_3$. Because
the diffusion equations are gauge-covariant, combinations that are
gauge-invariant on the boundary, such as
$O_H(x;t)=\Phi^{\dagger}(x;t)\Phi(x;t)$, remain gauge-invariant.

\subsection{One-loop renormalization}
\label{sec:oneg}
Provided the couplings and the fields of the theory have been
renormalized, any gauge-invariant operator remains UV finite at
positive flow times~\cite{Luscher:2011bx}. Even the wavefunction
renormalization of the gauge fields turns out to be
trivial~\cite{Luscher:2011bx}. However, the fermion fields undergo
wavefunction renormalization:
\begin{align}
  \psi(x;t)=Z_\psi^{1/2}(t)\psi_0(x),\quad
  \bar{\psi}(x;t)=Z_\psi^{1/2}(t)\bar{\psi}_0(x).
\end{align}
Using the result of~\cite{Luscher:2013cpa,Harlander:2018zpi}, we
compute the one-loop renormalization for $d=4-2\epsilon$ dimensions,
\begin{align}
  Z^{\flow}_\psi
  =1+\frac{1}{2(4\pi)^2}\frac{1}{\epsilon}
  \left[(6-2\xi)G_\psi+(2-\delta_L)y_\psi^2\right],
\end{align}
where
\begin{align}
  \label{eq:delta}
  G_\psi=\left(Y^2g^{\prime\,2}+\delta_LC_F^{(2)}g^2+\delta_CC_F^{(3)}g_s^2\right)
\end{align}
and the couplings are all in the $\MS$ scheme at the scale
$\mu^2=1/(2t)$.  $\xi$ is the gauge-fixing parameter ($\xi=0$ for
Landau gauge and $\xi=1$ for Feynman gauge) and
$C_F^{(N)}=(N^2-1)/(2N)$ are the quadratic Casimir eigenvalues of the
fundamental representation of $SU(N)$. We have assumed that the same
gauge condition is used for $U(1)_Y$ as for $SU(2)_L$. The values of
$\delta_L$, $\delta_C$ as well as $Y$ and the Yukawa coupling $y_\psi$
depend on the fermion. For left-handed fermions ($\delta_L=1$),
$|y_{\psi}|^2$ needs to be averaged over the doublet. For right-handed
fermions, it is multiplied by a factor of two since both $SU(2)_L$
components of the left handed fermion appear in the loop. These Higgs
loop contributions are only sizeable for the third quark
generation. Due to the unitarity of the CKM and PMNS matrices, the
above result is complete. For instance, starting from a right-handed
bottom quark, all the families of left-handed quarks can propagate and
we encounter the sum
$Y^{d*}_{3i}Y^d_{i3}=y_b^2V_{ib}^{\dagger}V^{~}_{bi}=y_b^2$. We could
also have argued in a coordinate-independent way, that $Y^d$ can be
diagonalized by independent left and right unitary transformations to
obtain the squared eigenvalue.  We remark that the renormalization
cannot depend on external parameters like the temperature. It is most
easily carried out using the Feynman rules of the original symmetric
Lagrangian.

The corresponding one-loop expression for the field in the $\MS$ scheme
($\psi^{\MS}(x;\mu)=[Z_\psi^{\MS}(\mu)]^{1/2}\psi_0(x)$) reads
\begin{align}
  Z^{\MS}_\psi=1-\frac{1}{2(4\pi)^2}\frac{1}{\epsilon}
  \left[2\xi\, G_{\psi}-(2-\delta_L)y_{\psi}^2\right].
\end{align}
The anomalous dimension function reads
\begin{align}
  \label{eq:gamma2}
  \gamma_\psi(\mu)=\frac{\dd Z_\psi(\mu)}{\dd\ln\mu^2}=\frac{1}{2(4\pi)^2}
  \left[2\xi\, G_{\psi}(\mu)-(2-\delta_L)y^2_\psi(\mu)\right].
\end{align}
The Yukawa couplings cancel from the gauge-invariant difference of the
$\gamma$-functions: due to the linear flow of both the scalar field
and the fermion fields, no new fermion-scalar vertices can be
generated in the bulk and only the boundary vertex needs to be
considered. The bulk propagation is free of new divergences and the
boundary divergence is the same in both schemes. Therefore, the
one-loop matching is obtained as
\begin{align}
  \psi(x;t)=\left\{ 1-\frac{3}{2(4\pi)^2}G_\psi(\mu)
    \left[\ln(2t\mu^2)-1\right]\right\}\psi^{\MS}(x;\mu).
\end{align}
This suggests to match the couplings at $\mu^2=1/(2t)$.

The renormalization of the coupling constants and their running with
$t$ is not needed for one-loop matching.  The conversion to the $\MS$
scheme for $\Phi$ is trivial at one-loop
order~\cite{Borgulat:2025gys}: $\phi(x;t)=\phi^{\MS}(x;\mu)$ with
$\mu^2\sim 1/t$.  For completeness, we list the one-loop
wavefunction renormalization for the scalar
field,
\begin{align}
  \label{eq:wave}
  Z_{\Phi}^{\flow}=Z_\Phi^{\MS}=1-\frac{\gamma_\Phi^{\MS}}{\epsilon},
\end{align}
where~\cite{Machacek:1983tz}
\begin{align}
  \label{eq:gamma}
  \gamma_{\Phi}^{\MS}=\frac{1}{(4\pi)^2}\left[(3-\xi)\left(g^{\prime\,2}+3g^2\right)-12y_t^2\right].
\end{align}
The pre-factors of $g^{\prime\,2}$, $g^2$ and $y_t$ are due to
$Y^2=1/4$, $C_F^{(2)}=3/4$ and $2N_C=6$, respectively. Above, we did
not write out the contributions from the smaller Yukawa couplings
$6y_b^2$, $6y_c^2$, $2y_{\tau}^2$, etc., which we will also omit below.

\subsection{Comment on the ringed scheme}
\label{sec:oner}
We start from the perturbative expansion
\begin{align}
  \left\langle\bar{\psi}(t)\overleftrightarrow{\slashed{D}}(t)\psi(t)\right\rangle=
  -\frac{(1+2\delta_C)(1+\delta_L)}{(4\pi t)^{2-\epsilon}}
  +\text{higher order}+\text{long distance},
\end{align}
where we used the Standard Model gauge group. The closed tadpole loop
gives the tree-level contribution, and the symmetric covariant
derivative is normalized according to eq.~\eqref{eq:symderiv}.  In QCD
the long-distance part is not visible in a naive perturbative
expansion but it contributes to the non-perturbative part of the trace
of the energy-momentum tensor. In the Standard Model one encounters
an additional $1/t$ correction, accompanied by
$\langle\Phi^\dagger\Phi\rangle$. In the Higgs phase, this will not be
strictly ``long distance'' anymore.

In~\cite{Makino:2014taa} it was suggested to use the above non-interacting
value as the renormalization condition:
\begin{align}
\mathring{Z}_\psi(t)
\left\langle\mathring{\bar{\psi}}(t)\overleftrightarrow{\slashed{D}}(t)\mathring{\psi}(t)\right\rangle\equiv
-\frac{(1+2\delta_C)(1+\delta_L)}{(4\pi t)^{2-\epsilon}}+\mathcal{O}(1/t),
\end{align}
where $\mathring{Z}_\psi=1+\text{singular}$. This normalization can
also be implemented in lattice regularization, dividing the fields by
the square root of the expectation value. Hopefully, at short flow
times, the non-perturbative correction can be neglected.  In this
ringed gradient-flow scheme~\cite{Makino:2014taa}:
\begin{align}
    \mathring{\psi}(x;t) &=\left\{1+\frac{3}{(4\pi)^2}G_\psi^{\MS}(\mu)
    \left[\ln(\mu^2 t)+\gamma_E-\ln 3-\frac{1}{3}\ln2\right]\right.\nonumber\\
    &\qquad\qquad\left.+\frac{y_\psi^2(2-\delta_L)}{2(4\pi)^2}
    \left[\ln(\mu^2 t)+\gamma_E+\ln 2-1\right]\right\}
    \psi^{\MS}(x;\mu).
\end{align}

This was generalized to scalar fields in~\cite{Borgulat:2025gys}:
\begin{align}
  \mathring{Z}_{\Phi}(t)\left\langle \mathring{\Phi}^\dagger(t)\mathring{\Phi}(t)\right\rangle=\frac{1}{(4\pi)^2t}+\text{finite},
\end{align}
where we have chosen the adequate normalization for the $SU(2)$ gauge
group. Then~\cite{Borgulat:2025gys}:
\begin{align}
  \label{eq:zeta}
  \mathring{\Phi}(x;t)=\left\{1-\frac{1}{(4\pi)^2}
    \left[(g^{\prime\,\MS})^2+3(g^{\MS})^2-\frac{y_t^2}{2}\right](1+2\ln 2)\right\}\Phi^{\MS}(x;\mu).
\end{align}

We already hinted above that this renormalization procedure introduces
some limitations in lattice regularization. Also from a purely
perturbative viewpoint there is a problem: in the Higgs phase of the
Standard Model, we have
\begin{align}
  \left\langle \Phi^\dagger(t)\Phi(t)\right\rangle=\frac{1}{(4\pi)^2t}+\text{higher order} + \text{finite},
\end{align}
where ``finite'' at leading order in $t$ is exactly
$\langle\rho^2(t)\rangle/2$, i.e.\ $\langle
O_H(t)\rangle_{\text{sub}}$, where the power-divergent contribution
proportional to the identity operator $\mathbbm{1}$ has been
subtracted. Moreover, in addition to the long-distance contribution
$(\langle\rho^2\rangle-v^2)/2$, this also contains the short-distance
contribution $v^2/2$. Therefore, this renormalization condition is not
compatible with a clean factorization within an OPE into short- and
long-distance contributions, and we will abstain from using the ringed
scheme.

\section{Connecting the gradient-flow to the \texorpdfstring{\mathversion{bold}$\MS$}{MS} scheme}
\label{sec:match}
The Standard Model is an EFT that is not defined for
$\mu\rightarrow\infty$ or, equivalently, at a flow time in the
vicinity of the $t=0$ boundary. However, this does not restrict the
applicability of perturbation theory or the perturbative matching
between schemes at scales much smaller than the cutoff. In order to
connect the symmetric theory with the theory that is formulated in
terms of the composite $SU(2)$-singlet fields for temperatures smaller
than the electroweak crossover temperature, $T_{\text{EWSB}}\approx
160\,$GeV~\cite{DOnofrio:2014rug,DOnofrio:2015gop,Gould:2022ran}, a
matching scale $\mu\gg m_t\approx y_t\,v/\sqrt{2}$,
e.g.\ $\mu=300\,$GeV, seems appropriate. The same holds for matching
the six-quark-flavour theory between the $\MS$ and the gradient-flow
schemes. We will work out an example for this matching, after discussing
the practical utility of the approach.

\subsection{Practical relevance of the gradient flow}
So far, our discussion on the necessity of formulating EWSB using
gradient-flow regularization has mostly focused on conceptual
issues. Of practical relevance is the definition of both the Higgs
vacuum expectation value and the physical Higgs particle. The mixing
between $O_H$ and the identity operator severely complicates this
definition and, alongside specific gauge-fixing choices, is at the
origin of large tadpole contributions and the poor convergence of
conventional perturbative series. Therefore, we will carry out the
matching of $O_H$ in detail in sections~\ref{sec:scal1}
and~\ref{sec:scal2} below.

A related issue is that at large momentum transfers $Q^2\gg v^2$,
subleading power corrections set in, making a systematic OPE with a
clean long- and short-distance factorization essential. The gradient
flow provides exactly this framework. The operators listed in
table~\ref{tab:operators} destroy the asymptotic physical states. The
short-distance interaction between the Higgs doublet and the $SU(2)_L$
vector bosons and fermions can be integrated out into short-distance
coefficient functions within an OPE. At very large $Q^2$,
beyond-leading-power corrections to Green functions become visible, as
discussed, for example, in~\cite{Maas:2020kda,Jenny:2022atm}, albeit
using the FMS formalism.

On a practical level, one can either set up amplitude calculations
within the standard $\MS$ scheme and track the operator mixing using
the classification achieved via the gradient flow, or one can carry
out the entire calculation natively using gradient-flow regularization
and convert the final result back into the $\MS$ scheme. The latter
approach ensures automatic UV-finiteness of all intermediate and final
matrix elements; however, the calculational tools are less well
developed.

In terms of the composite operators listed in
table~\ref{tab:operators}, the correspondence between $O_H$ and the
Higgs particle $|h\rangle$ is as follows:
\begin{align}
  \label{eq:41}
  \langle\Omega|O_H(0;t)|h(\vec{p})\rangle&=\left\langle\Omega\left|
  v(t)h(0;t)+\frac12h^2(0;t)\right|h(\vec{p})\right\rangle
  \nonumber\\
  &=\frac{v(t)}{\sqrt{2E_{\vec{p}}}}Z_h^{1/2}(t)+\frac12\langle\Omega|h^2(0;t)|h(\vec{p})\rangle ,
\end{align}
where the second term generates loop corrections $\sim\lambda
v\ln(tm_h^2)+\ldots$. Note that the unity operator does not contribute
to the above relation. Clearly, at tree level $O_H$ also creates pairs
of Higgs particles.

Up to power corrections, the remaining fundamental and composite
operators in table~\ref{tab:operators} correspond directly to the
on-shell particles that form the asymptotic states in scattering
experiments, since these are carried out in the low-temperature Higgs
phase. The matching of their on-shell properties to the $\MS$ scheme
is well established. Subsequently, these operators can be mapped
directly to the gradient-flow scheme at a finite flow time $t$ via
perturbative matching, allowing any long-distance matrix element in
the low-temperature phase to be rigorously evaluated using
experimental inputs.

\subsection{Matching of the scalar bilinear operator}
\label{sec:scal1}
The operator $O_H=\Phi^{\dagger}\Phi$ is not protected by any global
symmetry and will mix with the unity operator.  We start with the
flowed operator
\begin{align}
  \langle O_H(x;t)\rangle
  =\frac{c(t)}{t}\mathbbm{1}+\frac12\left\langle
  v^2(t)+2v(t)h(x;t)+h^2(x;t)\right\rangle.
\end{align}
The expectation value on the right-hand side appears to resemble a
long-distance matrix element. However, $\langle v^2(t)\rangle
=v^2(t)=m^2(t)/\lambda(t)$ can be expressed in terms of
short-distance couplings. Therefore, we factorize
\begin{align}
  \label{eq:ope1}
  2\left\langle O_H(t)\right\rangle =
  \frac{C_{0}(t)}{t}\mathbbm{1} + v^2(t) + \langle
  O_{\text{fluct}}(t)\rangle+ \mathcal{O}(t), \quad
\end{align}
where we have omitted the argument $x=0$ and defined
\begin{align}
  O_{\text{fluct}}(x;t)\coloneqq 2v(t)h(x;t)+h^2(x;t).
\end{align}
Note that subtracting the perturbative series
$C_{0}(t)/t$ will remove the power-term but introduce a
renormalon ambiguity in the definition of the long-distance
matrix element. Within eq.~\eqref{eq:41} and table~\ref{tab:operators},
$O_H$ can safely be replaced with $\frac12O_{\text{fluct}}$.

Composite local flowed operators with renormalized fields at $t > 0$
map continuously onto series of renormalized local boundary operators,
evaluated in the $\MS$ scheme at a renormalization scale $\mu$. For
the local scalar singlet operator:
\begin{align}
  \label{eq:ope2}
  \left\langle O_{\text{fluct}}(t)\right\rangle =
  \frac{1}{C_2(\mu,t)}
  \left[ C_{0,\MS}(\mu)m^2_{\OS} + \left\langle O_{\text{fluct},\MS}(\mu)\right\rangle
    + \mathcal{O}(m_{\OS}^{-2})\right],
\end{align}
where the scale-independent $m^2_{\OS}$ acts as the factorization
scale. We now have a renormalon ambiguity on both sides of the
equation.

The coefficient of the short-distance unity operator can be obtained as follows:
\begin{align}
  C_{0,\MS}(\mu)=\frac{2}{m^2_{\OS}}S\left\langle\Phi^\dagger_0(x)\Phi_0(x)\right\rangle_{\text{finite}},
\end{align}
where $S$ denotes the perturbative expansion of the expectation value
and $\Phi_0(x)$ is the bare boundary operator.

We can combine the above equations to define:
\begin{align}
  \label{eq:combine}
\left\langle O_{\text{fluct},\MS}(\mu)\right\rangle=C_2(\mu,t)\left\langle O_{\text{fluct}}(t)\right\rangle-\frac{1}{t}\left[C_2(\mu,t)C_0(t)
      - t\,m^2_{\OS}C_{0,\MS}(\mu)\right].
\end{align}
This provides a clean, finite, gauge-invariant definition of the
expectation value of the fluctuating field in the $\MS$ scheme. The
leading dimension-two renormalon~\cite{Beneke:1998ui} (at $u=1$ in the
Borel plane, see also~\cite{Bali:2014fea}) cancels between the two
terms in the square bracket. This is similar to the procedure
suggested to define the non-perturbative gluon condensate in QCD
in~\cite{Beneke:2025hlg}, see also~\cite{Harlander:2026lte}. One can
of course also invert this relation to determine $\langle
O_{\text{fluct}}(t)\rangle$ from $\langle
O_{\text{fluct},\MS}(\mu)\rangle$. Note that this expectation value
depends on $1/t$ in the gradient-flow scheme and logarithmically on
$\mu$ in the $\MS$ scheme.

\subsection{One-loop analysis of the matching coefficients}
\label{sec:scal2}
Using the standard loop integrals and group-theoretical factors, one
obtains the one-loop result in the $\MS$ scheme~\cite{Luo:2002ey}
\begin{align}
  \gamma_{m^2}^{\MS}=\frac{1}{4(4\pi)^2}\left[-24\lambda+3\left(g^{\prime\,2}+3g^2\right)-12y_t^2\right],
\end{align}
where we neglected the small Yukawa couplings.
Our normalization convention is (see eq.~\eqref{eq:gamma2})
\begin{align}
  \frac{\dd\ln Z_{m^2}(\mu)}{\dd\ln\mu^2}=
  -\frac{1}{m^2}\frac{\dd{}m^2}{\dd\ln\mu^2}=\gamma_{m^2}(\mu),
\end{align}
where the coefficient of the quartic Higgs term in the Lagrangian (see
eq.~\eqref{eq:higgs}) is accompanied by the coefficient $-\lambda$, not
$-\lambda/2$~\cite{Luo:2002ey} or
$-\lambda/4$~\cite{Borgulat:2025gys}. The anomalous dimension
$\gamma_{m^2}=-\gamma_{O_H}$ was already used in
section~\ref{sec:oner}, where we discussed the ringed renormalization
condition.

Using this, we can convert the $\MS$ quadratic Higgs coupling to the
on-shell scheme. There is some ambiguity in this procedure. Here
we compute $m_{\OS}$ using the background field
method~\cite{Abbott:1981ke} and the condition $p_\mu p^\mu=-m^2_{\OS}$:
\begin{align}
  \label{eq:oss}
  m^2_{\OS} &= m^2_{\MS}(\mu) \left\{ 1 +\frac{1}{4(4\pi)^2}\left[
    24\lambda(L-1)-3\left(g^{\prime\,2}+3g^2\right)\left(L-\frac{4}{3}\right)\right.\right.\nonumber\\
   &\qquad\qquad\qquad\left.\left. +12y_t^2\left(L-2\right)\right]\right\},\quad   L=\ln\left(\frac{m^2_{\OS}}{\mu^2}\right).
\end{align}

For the leading coefficient in the $\MS$ scheme, we obtain in the
non-interacting case
\begin{align}
  C_0^{\MS}(\mu)=\frac{1}{m^2_{\OS}}
  \frac{m^2(\mu)}{8\pi^2}\left[1- L\right].
\end{align}
The logarithm represents an infrared structure that is absorbed into
the long-distance matrix element within the OPE framework. Dropping
this logarithm and cancelling the mass-parameters using
eq.~\eqref{eq:oss} yields the short-distance result
\begin{align}
  C_0^{\MS}(\mu)=\frac{1}{8\pi^2}
\left\{ 1 -\frac{1}{4(4\pi)^2}\left[
  24\lambda(L-1)-3\left(g^{\prime\,2}+3g^2\right)\left(L-\frac{4}{3}\right)
 +12y_t^2\left(L-2\right)\right]\right\}.
\end{align}
On the gradient-flow side, we obtain
\begin{align}
  C_0(t)=\frac{1}{8\pi^2}\zeta^{-1}_{\Phi}
  =\frac{1}{8\pi^2}\left\{1+\frac{1}{(4\pi)^2}
  \left[g^{\prime\,2}+3g^2-\frac{y_t^2}{2}\right](1+2\ln 2)\right\},
\end{align}
where the conversion factor $\zeta_\Phi$ to the ringed scheme, in
which the free-field relation holds exactly, has been computed
in~\cite{Borgulat:2025gys} and can be read off eq.~\eqref{eq:zeta}.

The function $C_2$ absorbs the difference of the two wavefunction
renormalizations, which is absent at one-loop order, and also includes
the operator matching. Since the gradient-flowed operator does not
renormalize and $O_H^{\MS}(\mu)=[Z_{O_H}^{\MS}(\mu)]^{-1}O_{H,0}=
Z_{m^2}^{\MS}(\mu)O_{H,0}$, we obtain
\begin{align}
  C_2(\mu,t)&=1+
  \frac{1}{4(4\pi)^2}\left\{
  24\lambda\left[\ln(2t\mu^2)-1\right]
  -3(g^{\prime\,2}+3g^2)\left[\ln(2t\mu^2)-\frac{4}{3}\right]\right.\nonumber\\
  &\quad\quad\quad\left.
  +12y_t^2\left[\ln(2t\mu^2)-\frac{2}{3}\right]\right\}.
\end{align}
When these expressions are combined in the square bracket of
eq.~\eqref{eq:combine}, the result only depends on the combination
$tm^2_{\OS}$. As discussed above, a matching scale $t\ll m^{-2}_{\OS}$
is preferable.

\subsection[Matching of the composite and fundamental operators at order \texorpdfstring{$t$}{t}]{Matching of the composite and fundamental operators at order \texorpdfstring{\mathversion{bold}$t$}{t}}
Among the operators listed in table~\ref{tab:operators} only $O_H$
(and $v O_F$) mix with the identity.  All other operators do not have
the vacuum quantum numbers. The $\MS$-scheme operators are connected
to those at positive flow time via an OPE.

At small matching scales $\mu$ or, equivalently, large flow times the
composite (and fundamental) operators defining the physical low-energy
degrees of freedom receive subleading corrections of order $t$. These
become relevant at large momentum transfer $Q^2$. In this case,
$\mu^2\ll Q^2$ becomes a small scale. Below we list the
higher-dimensional operators that should be considered in the OPE.

We start from the OPE of the scalar operator that we already discussed above:
\begin{align}
  \left\langle O^{\MS}_H(\mu)\right\rangle = \frac{c_0(\mu,t)}{t}\mathbbm{1}
  +c_2(\mu,t)\left\langle O_H(t)\right\rangle_{\text{sub}}+t\sum_ic_{4,i}(\mu,t)\langle O_{H,4,i}(t)\rangle+\ldots.
\end{align}
Splitting $2O_H(x;t)$ into a short-distance part $v^2(t)$ and a long-distance
operator $O_{\text{fluct}}(x;t)=2O_H(x;t)-v^2(t)$, we obtain the
dimension-4 operators
\begin{align}
  O_{H,4}&\in\left\{ 
  v^2O_{\text{fluct}},\,
  O_{\text{fluct}}^2,\,
  F_{\mu\nu}F_{\mu\nu},\,
  Z^0_{\mu\nu}Z^0_{\mu\nu},\,
  W^+_{\mu\nu}W^-_{\mu\nu},\,
  G^A_{\mu\nu}G^A_{\mu\nu}\right.,\nonumber\\
&\qquad \left. vO_{F,ii}^{u,d,\nu,e},\, 
  \bar{\Psi}^{u,d,\nu,e}_{L,i}\overleftrightarrow{\slashed{D}}\Psi^{u,d,\nu,e}_{L,i},\,
  \bar{\Psi}^{u,d,\nu,e}_{R,i}\overleftrightarrow{\slashed{D}}\Psi^{u,d,\nu,e}_{R,i}\right\},
\end{align}
where $F_{\mu\nu}=\partial_{\mu}A_{\nu}-\partial_{\nu}A_{\mu}$,
$W^\pm_{\mu\nu}=\partial^{~}_\mu W^{\pm}_{\nu}-\partial^{~}_{\nu}W^{\pm}_\mu$
and $Z^0_{\mu\nu}=\partial^{~}_\mu Z^0_{\nu}-\partial^{~}_{\nu}Z^0_\mu$.  Note
that the covariant derivatives only contain the neutral gauge fields
and, if applied to quarks, $G_\mu$. The extra factor $v$ in front of
the mass-term operator $O_F$ appears because any loop correction
linking it to $O_H$ will be proportional to $v$. Also in terms of the
flow-time expansion, only integer powers of $t$ are possible.  The
first term is formally a short-distance correction to
$O_{\text{fluct}}$ itself, i.e.\ a power correction to $c_2(\mu,t)$.
In addition, there is a purely short-distance $tc_{4,v}v^4\mathbbm{1}$
contribution, i.e.\ an order-$t^2$ correction to $c_0(\mu,t)$. The
operators only contain the screened long-range fields (physical
basis), while the coefficient functions parameterize the
short-distance effects. We remark that on-shell the last three
operator classes are related by the equations of motion.

The composite fermions $\Psi_L$ and the fundamental fermions $\Psi_R=\psi_R$
both are $SU(2)_L$ gauge singlets. The operator list for $\Psi_L$ reads
\begin{align}
  O_{\Psi_L,4}&\in \left\{ v^2\Psi_L,\,
  O_{\text{fluct}}\Psi_L,\,\sigma_{\mu\nu}F_{\mu\nu}\Psi_L,\,
  \sigma_{\mu\nu}Z^0_{\mu\nu}\Psi_L,\,
  \sigma_{\mu\nu}G_{\mu\nu}^A t^A\Psi_L,\,\slashed{D}^2\Psi_L,
  v\slashed{D}\Psi_R\right\},
\end{align}
where the last two operators can be eliminated via the on-shell
equations of motion.  The gluon-field-strength term does not
contribute for leptons and the $F_{\mu\nu}$ term does not contribute
for neutrinos. The right-handed fermions have the same operator list
with $R\leftrightarrow L$ (which maps the mixed kinetic-mass term
to $v\slashed{D}\Psi_L$), and the anti-fermions receive analogous
contributions.

Finally, we list the operators that contribute to the composite vector
bosons at $\mathcal{O}(t)$:
\begin{align}
  O_{W^+_\mu,4} &\in \left\{
  v^2W^+_\mu ,\,
  O_{\text{fluct}} W^+_\mu ,\,
  \partial_\nu W_{\nu\mu}^+ ,\,
  \bar{\Psi}^{u,\nu}_{L,i} \gamma_\mu \Psi^{d,e}_{L,j}
  \right\},\\
  O_{W^-_\mu,4} &\in \left\{ 
  v^2W^-_\mu ,\,
  O_{\text{fluct}} W^-_\mu ,\,
  \partial_\nu W_{\nu\mu}^- ,\,
  \bar{\Psi}^{d,e}_{L,i} \gamma_\mu \Psi^{u,\nu}_{L,j}
  \right\},\\
  O_{\{Z^0_\mu, A_\mu\},4} &\in \left\{ 
  v^2Z^0_\mu ,\,
  v^2A_\mu ,\,
  O_{\text{fluct}} Z^0_\mu ,\,
  O_{\text{fluct}} A_\mu ,\,
  \partial_\nu Z^0_{\nu\mu} ,\,
  \partial_\nu F_{\nu\mu},\right. \nonumber\\
  &\left.\qquad\bar{\Psi}_{L,i} \gamma_\mu \Psi_{L,i} ,\,
  \bar{\Psi}_{R,i} \gamma_\mu \Psi_{R,i}
  \right\}.
\end{align}
Because the physical fields $A_\mu$ and $Z^0_\mu$ contain both the
$\mathbf{W}^3_\mu$ and $B_\mu$ structures, they share the same
operator list.

\section{Crossovers in the Standard Model and beyond}
\label{sec:cross}
As an aside, we remark that the proposed gradient-flow regularization
provides the natural mathematical framework to trace the thermal
evolution of the Standard Model. Because the flow time $t > 0$ acts as
a gauge- and Lorentz-invariant UV cutoff, it can track the running
parameters smoothly through both the electroweak and QCD crossovers,
connecting the low-energy observables cleanly back to the fundamental
short-distance fields. Before mapping out how the Standard Model fits
into the thermal history of the universe, we point out differences
between these two crossovers.

QCD with three massless quarks, $m_u=m_d=m_s=0$, possesses an exact
global $SU(3)_R\times SU(3)_L$ flavour symmetry. At low temperatures,
the $SU(3)_F$ chiral condensate $\frac13\langle
(O_{F,u}^u+O_{F,d}^d+O_{F,s}^d)\rangle$ (with
$O_{F,q}^{u,d}=\bar{\Psi}_q\Psi_q$ in the mass-diagonal basis)
spontaneously acquires a vacuum expectation value, and this symmetry
is broken to the vector $R+L$ $SU(3)_F$ symmetry, giving rise to eight
pseudoscalar Nambu-Goldstone bosons. As a result, the
quasi-particles --- namely, the hadrons --- of the low-temperature theory
are $SU(3)_C$ singlets. QCD with physical quark masses, which are
small compared to both $\Lambda_{\text{QCD}}$ and the chiral symmetry
breaking scale, inherits the non-vanishing chiral condensate and the
(now massive) pseudo-Goldstone bosons of the nearby $N_f=3$ massless
theory. However, the first-order phase transition of the massless
theory becomes a continuous crossover. In this low-temperature phase,
$SU(3)_C$ is confined, rendering the physical hadron spectrum strictly
colour-singlet.

In a complementary fashion, the local $SU(2)_L$ gauge symmetry of the
electroweak sector is screened such that its fundamental degrees of
freedom are effectively confined within the composite gauge-singlet
bosons and left-handed fermions. In the Standard Model no global
symmetry other than $U(1)_{B-L}$ exists. Unlike in QCD, no approximate
global symmetry exists either, exposing the standard textbook paradigm
of a custodial symmetry-breaking pattern as an artefact of specific
gauge-fixing choices. The defining feature that results in the
formation of composite gauge-singlet bosons and left-handed fermions
is the minimum of the Higgs potential along a critical $[SU(2)_L\times
  U(1)_Y]/U(1)_{\text{em}}$ gauge orbit. The vacuum condensate
$\langle O_H\rangle_{\text{sub}}$ plays a role similar to the chiral
condensate of QCD in triggering the long-distance screening of the
gauge interactions; however, its algebraic and field-theoretic origin
is fundamentally different. In~\cite{Francois:2017akk} the argument
was made that the only ``substantial'' gauge symmetry of the Standard
Model was $SU(3)\times U(1)_{\text{em}}$ since the remaining
``artificial'' gauge symmetry could be ``dressed'' with local
operators.  In contrast, screening the QCD and QED gauge symmetries
requires non-local dressing fields.  However, as discussed here, the
mixing of $\Phi^{\dagger}(x)\Phi(x)$ with the unity operator also
either requires a spatially non-local approach~\cite{Frohlich:1981yi}
or gradient-flow regularization, smearing out the operators over a
Euclidean distance $\sim\sqrt{8t}$.

We briefly track the thermal history of the universe:
\begin{itemize}
    \item The inflation crossover (or second order phase transition)
      and a strong first-order phase transition that triggered the
      abundance of matter over antimatter in the early universe, both
      at temperatures $T\gg T_{\text{EWSB}}$: these cannot be
      explained within the Standard Model.
    \item The electroweak crossover ($T \approx 160$\,GeV): as the
      temperature drops below $T_{\text{EWSB}} \approx
      160$\,GeV~\cite{DOnofrio:2014rug,Gould:2022ran}, the $SU(3)_C
      \times SU(2)_L \times U(1)_Y$ theory undergoes a smooth,
      analytic crossover. No local gauge symmetry is broken; instead,
      the physical degrees of freedom are rearranged into new
      quasi-particle bound states. The gauge-invariant scalar singlet
      density $\langle \Phi^\dagger\Phi \rangle$ undergoes a sudden,
      rapid increase towards its low-temperature vacuum expectation
      value. The left-handed fermions and longitudinal gauge fields
      screen the weak interactions, leaving only the $SU(3)_C \times
      U(1)_{\text{em}}$ gauge combination to survive as the
      long-distance degrees of freedom.
    \item The QCD crossover ($T \approx 155$\,MeV): as the
      temperature drops below $T_{\text{QCD}} \approx
      155$\,MeV~\cite{Borsanyi:2010bp,HotQCD:2018pds}, a second
      analytic crossover binds the quarks and gluons into
      colour-singlet hadrons. The $SU(3)_C$ colour charge becomes
      completely confined at long distances, leaving
      $U(1)_{\text{em}}$ electromagnetism as the sole unscreened,
      long-range gauge interaction.  Most of the baryonic mass is
      generated at this crossover; only a small fraction, represented
      by the light $\sigma$-terms, owes its origin to EWSB.
    \item The cosmic recombination crossover ($T_{\text{dec}}\approx
      0.3\,$eV): nuclei and electrons combine into atoms, liberating
      the photons. This is the origin of the cosmic microwave
      background. The decoupling temperature $T_{\text{dec}}$ is much
      lower than the ionization energy of hydrogen, due to the large
      photon density in the early
      universe~\cite{Peebles:1968ja,Zeldovich:1969ff}. The term
      ``recombination'' is a historical misnomer. In the decoupled
      phase, stable atoms form and electric charges are screened.
\end{itemize}
In spite of the above, even at $T=0$ residual strong forces exist
between hadrons (nuclear physics) and quarks and gluons can be
liberated in deep-inelastic scattering. Atoms can still combine into
molecules (chemistry) and there are van der Waals interactions between
molecules. Moreover, electric charges are not entirely screened at the
energy transfers, distances and localized temperatures that are
encountered in our natural environment.

The above screening trajectory exposes a profound irony regarding the
UV-completeness of the theory. The $SU(3)_C$ and $SU(2)_L$ sectors are
natively UV-complete due to asymptotic freedom. Modern renormalization
group
analyses~\cite{Bezrukov:2012sa,Degrassi:2012ry,Buttazzo:2013uya,Bednyakov:2015sca,Bednyakov:2025uur}
suggest that the Higgs sector may also approach a pseudo-Gaussian
fixed point (asymptotic criticality) in the UV. However, absolute UV
completion is obstructed by the running of the $U(1)_Y$ and Yukawa
couplings, though their respective Landau poles lie well beyond the
Planck scale.

Essentially, asymptotic safety is spoiled by the only gauge symmetry
that survives unscreened at long distances, $U(1)_{\text{em}}$, whose
coupling --- unless embedded in a non-Abelian theory --- will diverge at
very short distances, albeit far beyond the Planck scale. At the same
time, this is precisely the physical mechanism that prevents the
universe from becoming entirely sterile. Because the fine-structure
coupling $e$ runs to small values at low energies, electromagnetism
escapes the screening trap that neutralizes the strong and weak forces
at long distances. If nature were populated exclusively by
UV-complete, asymptotically free gauge theories, cooling the system
would force all long-range degrees of freedom to rearrange into
localized, massive bound states. It is precisely because the $U(1)$
and gravitational couplings grow at high energies and decrease at low
energies that long-range force fields can survive at low temperatures
to build chemical and biological matter, alongside large-scale cosmic
structures.

\section{Summary}
We have presented a description of the Standard Model Higgs mechanism
in a manifestly gauge-covariant framework. Employing the gradient flow
provides a structurally clean formalism to define the vacuum
expectation value of the Higgs condensate in the low-temperature phase
of the universe. This vacuum expectation value is finite and remains
well-separated from the unity operator, as detailed in
section~\ref{sec:scal1}.  In this formulation, the rearrangement of
the fundamental fields into the physical basis fields relevant at low
temperatures --- including the physical Higgs boson --- becomes highly
transparent.  While previous gauge-invariant treatments of electroweak
``symmetry breaking'' relied on spatially non-local operators that
complicated practical calculations, the local approach developed here
offers distinct practical advantages. For instance, we demonstrated
how the gradient flow eliminates the leading dimension-two renormalon
ambiguity of the Higgs condensate.

Within particle physics, the gradient flow has primarily received
attention in the context of lattice QCD computations, and its
perturbative toolkit remains less developed than that of modified
minimal subtraction. However, the scope of gradient-flow
regularization has steadily broadened; for example,
in~\cite{Crosas:2026ofx} the matching of the low-energy EFT (LEFT) of
the Standard Model EFT (SMEFT) to QCD in the gradient-flow scheme was
carried out. The electroweak sector of the Standard Model, which at
zero temperature does not require non-perturbative methods, represents
another highly compelling avenue for further gradient-flow
applications.

\providecommand{\href}[2]{#2}\begingroup\raggedright\endgroup
\end{document}